# A Scalable Algorithm for Maximizing Range Sum in Spatial Databases

Dong-Wan Choi[1]　　　Chin-Wan Chung[1,2]　　　Yufei Tao[2,3]

[1]Department of Computer Science, KAIST, Daejeon, Korea
[2]Division of Web Science and Technology, KAIST, Daejeon, Korea
[3]Department of Computer Science and Engineering,
Chinese University of Hong Kong, New Territories, Hong Kong

dongwan@islab.kaist.ac.kr　　　chungcw@kaist.edu　　　taoyf@cse.cuhk.edu.hk

## ABSTRACT

This paper investigates the *MaxRS* problem in spatial databases. Given a set $O$ of weighted points and a rectangular region $r$ of a given size, the goal of the MaxRS problem is to find a location of $r$ such that the sum of the weights of all the points covered by $r$ is maximized. This problem is useful in many location-based applications such as finding the best place for a new franchise store with a limited delivery range and finding the most attractive place for a tourist with a limited reachable range. However, the problem has been studied mainly in theory, particularly, in computational geometry. The existing algorithms from the computational geometry community are in-memory algorithms which do not guarantee the scalability. In this paper, we propose a scalable external-memory algorithm (*ExactMaxRS*) for the MaxRS problem, which is optimal in terms of the I/O complexity. Furthermore, we propose an approximation algorithm (*ApproxMaxCRS*) for the *MaxCRS* problem that is a circle version of the MaxRS problem. We prove the correctness and optimality of the ExactMaxRS algorithm along with the approximation bound of the ApproxMaxCRS algorithm. From extensive experimental results, we show that the ExactMaxRS algorithm is *two orders of magnitude* faster than methods adapted from existing algorithms, and the approximation bound in practice is much better than the theoretical bound of the ApproxMaxCRS algorithm.

## 1. INTRODUCTION

In the era of mobile devices, *location-based services* are being used in a variety of contexts such as emergency, navigation, and tour planning. Essentially, these applications require managing and processing a large amount of location information, and technologies studied in spatial databases are getting a great deal of attention for this purpose. Traditional researches in spatial databases, however, have mostly focused on retrieving objects (e.g., *range search*, *nearest neighbor search*, etc.), rather than finding the best location to optimize a certain objective.

Recently, several *location selection problems* [9, 16, 18, 19, 20, 21, 22, 23] have been proposed. One type of these problems is to find a location for a new facility by applying the well-known *facility location problem* in theory to database problems such as *optimal-location queries* and *bichromatic reverse nearest neighbor queries*. Another type of location selection problems is to choose one of the predefined candidate locations based on a given ranking function such as *spatial preference queries*.

In this paper, we solve the *maximizing range sum (MaxRS) problem* in spatial databases. Given a set $O$ of weighted points (a.k.a. objects) and a rectangle $r$ of a given size, the goal of the MaxRS problem is to find a location of $r$ which maximizes the sum of the weights of all the objects covered by $r$. Figure 1 shows an instance of the MaxRS problem where the size of $r$ is specified as $d_1 \times d_2$. In this example, if we assume that the weights of all the objects are equally set to 1, the center point of the rectangle in solid line is the solution, since it covers the largest number of objects which is 8. The figure also shows some other positions for $r$, but it should be noted that there are *infinitely many* such positions – $r$ can be anywhere in the data space. The MaxRS problem is different from existing location selection problems mentioned earlier in that there are no predefined candidate locations or other facilities to compete with. Furthermore, this problem is also different from *range aggregate queries* [17] in the sense that we do not have a known query rectangle, but rather, must discover the best rectangle in the data space.

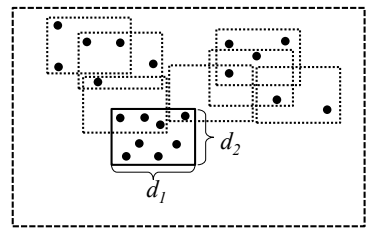

**Figure 1: An instance of the MaxRS problem**

In practice, there can be many kinds of facilities that should be associated with a region of a certain size. For example, if we open, in an area with a grid shaped road network, a new pizza franchise store that has a limited delivery range, it is important to maximize the number of residents in a rectangular area around the pizza store. This is the case of finding a more profitable place to set up a new service facility.

For an opposite case, the MaxRS problem can be applied to find a more serviceable place for client users. Consider a tourist who wants to find the most representative spot in a city. In this case, the tourist will prefer to visit as many attractions as possible around the spot, and at the same time s/he usually does not want to go too far away from the spot.

There has been little research for this natural problem in the database community. In fact, this problem has been mainly studied in the computational geometry community. The first optimal in-memory algorithm for finding the position of a fixed-size rectangle enclosing the maximum number of points was proposed in [11]. Later, a solution to the problem of finding the position of a fixed-size circle enclosing the maximum number of points was provided in [4].

Unfortunately, these in-memory algorithms are not scalable for processing a large number of geographic objects in real applications such as residential buildings and mobile customers, since they are developed based on the assumption that the entire dataset can be loaded in the main memory. A straightforward adaptation of these in-memory algorithms into the external memory can be considerably inefficient due to the occurrence of excessive I/O's.

In this paper, we propose the first external-memory algorithm, called *ExactMaxRS*, for the maximizing range sum (MaxRS) problem. The basic processing scheme of ExactMaxRS follows the *distribution-sweep* paradigm [10], which was introduced as an external version of the *plane-sweep* algorithm. Basically, we divide the entire dataset into smaller sets, and recursively process the smaller datasets until the size of a dataset gets small enough to fit in memory. By doing this, the ExactMaxRS algorithm gives an exact solution to the MaxRS problem. We derive the upper bound of the I/O complexity of the algorithm. Indeed, this upper bound is proved to be the lower bound under the comparison model in external memory, which implies that our algorithm is optimal.

Furthermore, we propose an approximation algorithm, called *ApproxMaxCRS*, for the *maximizing circular range sum (MaxCRS) problem*. This problem is the circle version of the MaxRS problem, and is more useful than the rectangle version, when a boundary with the same distance from a location is required. In order to solve the MaxCRS problem, we apply the ExactMaxRS algorithm to the set of Minimum Bounding Rectangles (MBR) of the data circles. After obtaining a solution from the ExactMaxRS algorithm, we find an approximate solution for the MaxCRS problem by choosing one of the candidate points, which are generated from the point returned from the ExactMaxRS algorithm. We prove that ApproxMaxCRS gives a (1/4)-approximate solution in the worst case, and also show by experiments that the approximation ratio is much better in practice.

**Contributions.** We summarize our main contributions as follows:

- We propose the ExactMaxRS algorithm, the first external-memory algorithm for the MaxRS problem. We also prove both the correctness and optimality of the algorithm.

- We propose the ApproxMaxCRS algorithm, an approximation algorithm for the MaxCRS problem. We also prove the correctness as well as tightness of the approximation bound with regard to this algorithm.

- We experimentally evaluate our algorithms using both real and synthetic datasets. From the experimental results, we show that the ExactMaxRS algorithm is two orders of magnitude faster than methods adapted from existing algorithms, and the approximation bound of the ApproxMaxCRS algorithm in practice is much better than its theoretical bound.

**Organization.** In Section 2, we formally define the problems studied in this paper, and explain our computation model. In Section 3, related work is discussed. In Section 4, we review the in-memory algorithms proposed in the computational geometry community. In Sections 5 and 6, the ExactMaxRS algorithm and ApproxMaxCRS algorithm are derived, respectively. In Section 7, we show experimental results. Conclusions are made and future work is discussed in Section 8.

## 2. PROBLEM FORMULATION

Let us consider a set of spatial objects, denoted by $O$. Each object $o \in O$ is located at a point in the 2-dimensional space, and has a non-negative *weight* $w(o)$. We also use $P$ to denote the infinite set of points in the entire data space.

Let $r(p)$ be a rectangular region of a given size centered at a point $p \in P$, and $O_{r(p)}$ be the set of objects covered by $r(p)$. Then the maximizing range sum (MaxRS) problem is formally defined as follows:

DEFINITION 1 (MAXRS PROBLEM). *Given $P$, $O$, and a rectangle of a given size, find a location $p$ that maximizes:*

$$\sum_{o \in O_{r(p)}} w(o).$$

Similarly, let $c(p)$ be a circular region centered at $p$ with a given diameter, and $O_{c(p)}$ be the set of objects covered by $c(p)$. Then we define the maximizing circular range sum (MaxCRS) problem as follows:

DEFINITION 2 (MAXCRS PROBLEM). *Given $P$, $O$, and a circle of a given diameter, find a location $p$ that maximizes:*

$$\sum_{o \in O_{c(p)}} w(o).$$

For simplicity, we discuss only the SUM function in this paper, even though our algorithms can be applied to other aggregates such as COUNT, SUM, and AVERAGE. Without loss of generality, objects on the boundary of the rectangle or the circle are excluded.

Since we focus on a massive number of objects that do not fit in the main memory, the whole dataset $O$ is assumed to be stored in external memory such as a disk. Therefore, we follow the standard external memory (EM) model [10] to develop and analyze our algorithms. According to the EM model, we use the following parameters:

$N$ : the number of objects in the database (i.e., $|O|$)
$M$ : the number of objects that can fit in the main memory
$B$ : the number of objects per block

We comply with the assumption that $N$ is much larger than $M$ and $B$, and the main memory has at least two blocks (i.e., $M \geq 2B$).

In the EM model, the time of an algorithm is measured by the number of I/O's rather than the number of basic operations as in the random access memory (RAM) model. Thus, when we say linear time in the EM model, it means that the number of blocks transferred between the disk and memory is bounded by $O(N/B)$ instead of $O(N)$. Our goal is to minimize the total number of I/O's in our algorithms.

## 3. RELATED WORK

We first review the range aggregate processing methods in spatial databases. The range aggregate (RA) query was proposed for the scenario where users are interested in summarized information about objects in a given range rather than individual objects. Thus, a RA query returns an aggregation value over objects qualified for a given range. In order to efficiently process RA queries, usually *aggregate indexes* [5, 12, 13, 15, 17] are deployed as the underlying access method. To calculate the aggregate value of a query region, a common idea is to store a pre-calculated value for each entry in the index, which usually indicates the aggregation of the region specified by the entry. However, the MaxRS problem cannot be efficiently solved using aggregate indexes, because the key is to find out *where* the best rectangle is. A naive solution to the MaxRS problem is to issue an infinite number of RA queries, which is prohibitively expensive.

Recently, researches about the selection of optimal locations in spatial databases have been reported, and they are the previous work most related to ours. Du et al. proposed the *optimal-location query* [9], which returns a location in a query region to maximize the *influence* that is defined to be the total weight of the reverse nearest neighbors. They also defined a different query semantics in their extension [22], called *min-dist optimal-location query*. In both works, their problems are stated under $L_1$ distance. Similarly, the *maximizing bichromatic nearest neighbor (MaxBRNN) problem* was studied by Wong et al. [18] and Zhou et al. [23]. This is similar to the problem in [9] except that $L_2$ distance, instead of $L_1$ distance, is considered, making the problem more difficult. Moreover, Xiao et al. [20] applied optimal-location queries to road network environments.

However, all these works share the spirit of the classic facility location problem, where there are two kinds of objects such as customers and service sites. The goal of these works is essentially to find a location that is far from the competitors and yet close to customers. This is different from the MaxRS (MaxCRS) problem, since we aim at finding a location with the maximum number of objects around, without considering any competitors. We have seen the usefulness of this configuration in Section 1.

There is another type of location selection problems, where the goal is to find top-k spatial sites based on a given ranking function such as the weight of the nearest neighbor. Xia et al. proposed the *top-t most influential site query* [19]. Later, the *top-k spatial preference query* was proposed in [16, 21], which deals with a set of classified feature objects such as hotels, restaurants, and markets by extending the previous work. Even though some of these works consider the range sum function as a ranking function, their goal is to choose one of the candidate locations that are predefined. However, there are an infinite number of candidate locations in the MaxRS (MaxCRS) problem, which implies that these algorithms are not applicable to the problem we are focusing on.

In the theoretical perspective, MaxRS and MaxCRS have been studied in the past. Specifically, in the computational geometry community, there were active researches for the *max-enclosing polygon problem*. The purpose is to find a position of a given polygon to enclose the maximum number of points. This is almost the same as the MaxRS problem, when a polygon is a rectangle. For the *max-enclosing rectangle problem*, Imai et al. proposed an optimal in-memory algorithm [11] whose time complexity is $O(n \log n)$, where $n$ is the number of rectangles. Actually, they solved a problem of finding the maximum clique in the rectangle intersection graph based on the well-known plane-sweep algorithm, which can be also used to solve the max-enclosing rectangle problem by means of a simple transformation [14]. Inherently, however, these in-memory algorithms do not consider a scalable environment that we are focusing on.

In company with the above works, there were also works to solve the *max-enclosing circle problem*, which is similar to the MaxCRS problem. Chazelle et al. [4] were the first to propose an $O(n^2)$ algorithm for this problem by finding a maximum clique in a circle intersection graph. The max-enclosing circle problem is actually known to be $3$SUM-*hard* [3], namely, it is widely conjectured that no algorithm can terminate in less than $\Omega(n^2)$ time in the worst case. Therefore, several approximation approaches were proposed to reduce the time complexity. Recently, Berg et al. proposed a $(1 - \epsilon)$-approximation algorithm [7] with time complexity $O(n \log n + n\epsilon^{-3})$. They divide the entire dataset into a grid, and then compute the local optimal solution for a grid cell. After that the local solutions of cells are combined using a dynamic-programming scheme. However, it is generally known that a standard implementation of dynamic programming leads to poor I/O performance [6], which is the reason why it is difficult for this algorithm to be scalable.

## 4. PRELIMINARIES

In this section, we explain more details about the solutions proposed in the computational geometry community. Our solution also shares some of the ideas behind those works. In addition, we show that the existing solutions cannot be easily adapted to our environment, where a massive size of data is considered.

First, let us review the idea of transforming the *max-enclosing rectangle problem* into the *rectangle intersection problem* in [14]. The max-enclosing rectangle problem is the same as the MaxRS problem except that it considers only the count of the objects covered by a rectangle (equivalently, each object has weight 1). The rectangle intersection problem is defined as *"Given a set of rectangles, find an area where the most rectangles intersect"*. Even though these two problems appear to be different at first glance, it has been proved that the max-enclosing rectangle problem can be mapped to the rectangle intersection problem [14].

We explain this by introducing a mapping example shown in Figure 2. Suppose that the dataset has four objects

(black-filled) as shown in Figure 2(a). Given a rectangle of size $d_1 \times d_2$, an optimal point can be the center point $p$ of rectangle $r$ (see Figure 2(a)). To transform the problem, we draw a rectangle of the same size centered at the location of each object as shown in Figure 2(b). It is not difficult to observe that the optimal point $p$ in the max-enclosing rectangle problem can be any point in the most overlapped area (gray-filled) which is the outcome of the rectangle intersection problem. Thus, once we have found the most overlapped area in the transformed rectangle intersection problem, the optimal location of the max-enclosing rectangle problem can trivially be obtained.

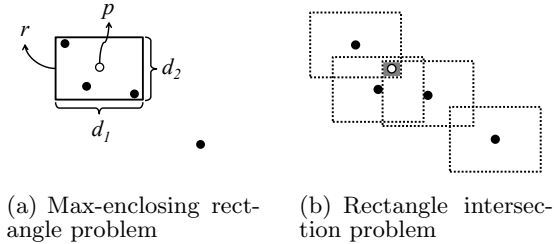

(a) Max-enclosing rectangle problem  (b) Rectangle intersection problem

**Figure 2: An example of transformation**

For the rectangle intersection problem, an *in-memory* algorithm was proposed in [11], which is based on the well-known *plane-sweep* algorithm. Basically, the algorithm regards the edges of rectangles as intervals and maintains a binary tree while sweeping a conceptual horizontal line from bottom to top. When the line meets the bottom (top) edge of a rectangle, a corresponding interval is inserted to (deleted from) the binary tree, along with updating the *count*s of intervals currently residing in the tree, where the count of an interval indicates the number of intersecting rectangles within the interval. An interval with the maximum count during the whole sweeping process is returned as the final result. The time complexity of this algorithm is $O(n \log n)$, where $n$ is the number of rectangles, since $n$ insertions and $n$ deletions are performed during the sweep, and the cost of each tree operation is $O(\log n)$. This is the best efficiency possible in terms of the number of comparisons [11].

Unfortunately, this algorithm cannot be directly applied to our environment that is focused on massive datasets, since the plane-sweep algorithm is an in-memory algorithm based on the RAM model. Furthermore, a straightforward adaptation of using the B-tree instead of the binary tree still requires a large amount of I/O's, in fact $O(N \log_B N)$. Note that the factor of $N$ is very expensive in the sense that linear cost is only $O(N/B)$ in the EM model.

## 5. EXACT ALGORITHM FOR MAXIMIZING RANGE SUM

In this section, we propose an external-memory algorithm, namely *ExactMaxRS*, that exactly solves the MaxRS problem in $O((N/B) \log_{M/B} (N/B))$ I/O's. This is known [2, 11] to be the lower bound under the comparison model in external memory.

### 5.1 Overview

Essentially, our solution is based upon the transformation explained in Section 4. Specifically, to transform the MaxRS problem, for each object $o \in O$, we construct a corresponding rectangle $r_o$ which is centered at the location of $o$ and has a weight $w(o)$. All these rectangles have the same size, which is as specified in the original problem. We use $R$ to denote the set of these rectangles. Also, we define two notions which are needed to define our transformed MaxRS problem later:

DEFINITION 3 (LOCATION-WEIGHT). *Let $p$ be a location in $P$, the infinite set of points in the entire data space. Its location-weight with regard to $R$ equals the sum of the weights of all the rectangles in $R$ that cover $p$.*

DEFINITION 4 (MAX-REGION). *The max-region $\rho$ with regard to $R$ is a rectangle such that:*

- *every point in $\rho$ has the same location-weight $\tau$, and*
- *no point in the data space has a location-weight higher than $\tau$.*

Intuitively, the max-region $\rho$ with regard to $R$ is an intersecting region with the maximum *sum* of the weights of the overlapping rectangles. Then our transformed MaxRS problem can be defined as follows:

DEFINITION 5 (TRANSFORMED MAXRS PROBLEM). *Given $R$, find a max-region $\rho$ with regard to $R$.*

Apparently, once the above problem is solved, we can return an arbitrary point in $\rho$ as the answer for the original MaxRS problem.

At a high level, the ExactMaxRS algorithm follows the divide-and-conquer strategy, where the entire datset is recursively divided into mutually disjoint subsets, and then the solutions that are locally obtained in the subsets are combined. The overall process of the ExactMaxRS algorithm is as follows:

1. Recursively divide the whole space vertically into $m$ sub-spaces, called *slab*s and denoted as $\gamma_1, ,, \gamma_m$, each of which contains roughly the same number of rectangles, until the rectangles belonging to each slab can fit in the main memory.

2. Compute a solution structure for each slab, called *slab-file*, which represents the local solution to the sub-problem with regard to the slab.

3. Merge $m$ slab-files to compute the slab-file for the union of the $m$ slabs until the only one slab-file remains.

In this process, we need to consider the following: (1) How to divide the space to guarantee the termination of recursion; (2) how to organize slab-files, and what should be included in a slab-file; (3) how to merge the slab-files without loss of any necessary information for finding the final solution.

### 5.2 ExactMaxRS

Next we address each of the above considerations, and explain in detail our ExactMaxRS algorithm.

#### 5.2.1 Division Phase

Let us start with describing our method for dividing the entire space. Basically, we recursively divide the space vertically into $m$ slabs along the x-dimension until the number of rectangles in a slab can fit in the main memory. Since a rectangle in $R$ can be large, it is unavoidable that a rectangle may need to be split into a set of smaller disjoint rectangles as the recursion progresses, which is shown in Figure 3. As a naive approach, we could just insert all the split rectangles into the corresponding slabs at the next level of recursion. In Figure 3, the three parts of rectangle $r$ will be inserted into slabs $\gamma_1$, $\gamma_2$, and $\gamma_3$, respectively.

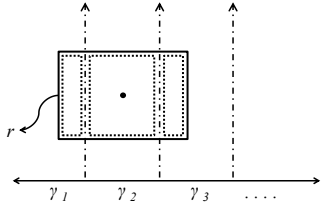

Figure 3: An example of splitting a rectangle

However, it is not hard to see that this approach does not guarantee the termination of recursion, since rectangles may *span* an entire slab, e.g., the middle part of $r$ spans slab $\gamma_2$. In the extreme case, suppose that all rectangles span a slab $\gamma$. Thus, no matter how many times we divide $\gamma$ into sub-slabs, the number of rectangles in each sub-slab still remains the same, meaning that recursion will never terminate infinitely.

Therefore, in order to gradually reduce the number of rectangles for each sub-problem, we do not pass spanning rectangles to the next level of recursion, e.g., the middle part of $r$ will not be inserted in the input of the sub-problem with regard to $\gamma_2$. Instead, the spanning rectangles are considered as another local solution for a separate, special, sub-problem. Thus, in the merging phase, the spanning rectangles are also merged along with the other slab-files. In this way, it is guaranteed that recursion will terminate eventually as proved in the following lemma:

LEMMA 1. *After $O(\log_m(N/M))$ recursion steps, the number of rectangles in each slab will fit in the main memory.*

PROOF. Since the spanning rectangles do not flow down to the next recursion step, we can just partition the vertical edges of rectangles. There are initially $2N$ vertical edges. The number of edges in a sub-problem will be reduced by a factor of $m$ by dividing the set of edges into $m$ smaller sets each of which has roughly the same size. Each vertical edge in a slab represents a split rectangle. It is obvious that there exists an $h$ such that $2N/m^h \leq M$. The smallest such $h$ is thus $O(\log_m(N/M))$. □

**Determination of $m$.** We set $m = \Theta(M/B)$, where $M/B$ is the number of blocks in the main memory.

### 5.2.2 Slab-files

The next important question is how to organize a slab-file. What the question truly asks about is what structure should be returned after *conquering* the sub-problem with regard to a slab. Each slab-file should have enough information to find the final solution after all the merging phases.

To get the intuition behind our solution (to be clarified shortly), let us first consider an easy scenario where every rectangle has weight 1, and is small enough to be totally inside a slab, which is shown in Figure 4. Thus, no spanning rectangle exists. In this case, all we have to do is to just maintain a max-region (black-filled in Figure 4) with regard to rectangles in each slab. Recall that a max-region is the most overlapped area with respect to the rectangles in the corresponding slab (see Definition 4). Then, in the merging phase, among $m$ max-regions (i.e., one for each slab), we can choose the best one as the final solution. In Figure 4, for instance, the best one is $\rho_m$ because it is the intersection of 3 rectangles, whereas the number is 2 for the max regions of the other slabs.

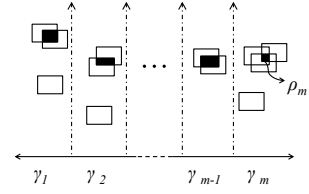

Figure 4: An easy scenario to illustrate the intuition of slab-files

Extending the above idea, we further observe that the horizontal boundaries of a max-region are laid on the horizontal lines passing the bottom or top edge of a certain rectangle. Let us use the term *h-line* to refer to a horizontal line passing a horizontal edge of an input rectangle. Therefore, for each h-line in a slab, it suffices to maintain a segment that could belong to the max-region of the slab. To formalize this intuition, we define *max-interval* as follows:

DEFINITION 6 (MAX-INTERVAL). *Let (1) $\ell.y$ be the y-coordinate of a h-line $\ell$, and $\ell_1$ and $\ell_2$ be the consecutive h-lines such that $\ell_1.y < \ell_2.y$, (2) $\ell \cap \gamma$ be the part of a h-line $\ell$ in a slab $\gamma$, and (3) $r_\gamma$ be the rectangle formed by $\ell_1.y$, $\ell_2.y$, and vertical boundaries of $\gamma$. A max-interval is a segment $t$ on $\ell_1 \cap \gamma$ such that, the x-range of $t$ is the x-range of the rectangle $r_{max}$ bounded by $\ell_1.y$, $\ell_2.y$, and vertical lines at $x_i$ and $x_j$, where each point in $r_{max}$ has the maximum location-weight among all points in $r_\gamma$.*

Figure 5 illustrates Definition 6.

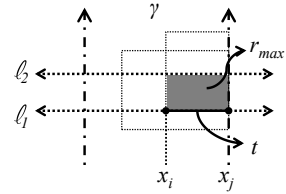

Figure 5: An illustration of Definition 6

Our slab-file is a set of max-intervals defined *only on h-lines*. Specifically, each max-interval is represented as a tuple specified as follows:

$$t = < y, \ [x_1, x_2], \ sum >$$

where $y$ is the y-coordinate of $t$ (hence, also of the h-line that defines it), and $[x_1, x_2]$ is the x-range of $t$, and $sum$ is the location-weight of any point in $t$. In addition, all the tuples in a slab-file should be sorted in ascending order of y-coordinates.

*Example 1.* Figure 6 shows the slab-files that are generated from the example in Figure 2, assuming that $m = 4$ and $\forall o \in O$, $w(o) = 1$. Max-intervals are represented as solid segments. For instance, the slab-file of slab $\gamma_1$ consists of tuples (in this order): $< y_2, [x_1, x_2], 1 >$, $< y_4, [x_1, x_2], 2 >$, $< y_6, [x_0, x_2], 1 >$, $< y_7, [-\infty, x_2], 0 >$. The first tuple $< y_2, [x_1, x_2], 1 >$ implies that, in slab $\gamma_1$, on any horizontal line with y-coordinate in $(y_2, y_4)$, the max-interval is always $[x_1, x_2]$, and its *sum* is 1. Similarly, the second tuple $< y_4, [x_1, x_2], 2 >$ indicates that, on any horizontal line with y-coordinate in $(y_4, y_6)$, $[x_1, x_2]$ is always the max-interval, and its *sum* is 2. Note that spanning rectangles have not been counted yet in these slab-files, since (as mentioned earlier) they are not part of the input to the sub-problems with regard to slabs $\gamma_1, ..., \gamma_4$.

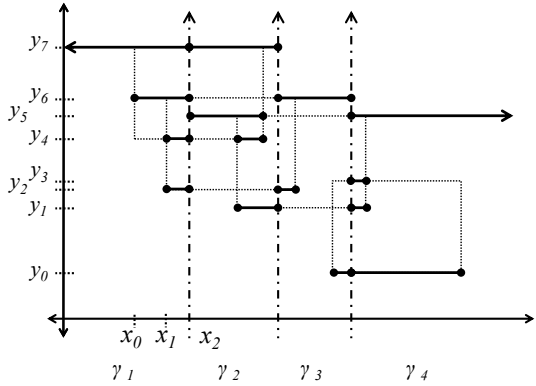

**Figure 6: An example of slab-files**

LEMMA 2. *Let $K$ be the number of rectangles in a slab. Then the number of tuples in the corresponding slab-file is $O(K)$.*

PROOF. The number of h-lines is at most double the number of rectangles. As a h-line defines only one max-interval in each slab, the number of tuples in a slab-file is at most $2K$, which is $O(K)$. □

### 5.2.3 Merging Phase

Now we tackle the last challenge: how to merge the slab-files, which is also the main part of our algorithm.

The merging phase sweeps a horizontal line across the slab-files and the file containing spanning rectangles. At each h-line, we choose a max-interval with the greatest *sum* among the max-intervals with regard to the $m$ slabs, respectively. Sometimes, max-intervals from adjacent slabs are combined into a longer max-interval.

The details of merging, namely *MergeSweep*, are presented in Algorithm 1. The input includes a set of spanning rectangles and $m$ slab-files. Also, each spanning rectangle contains only the spanning part cropped out of the original rectangle $r_o \in R$, and has the same weight as $r_o$ (recall that the weight of $r_o$ is set to $w(o)$). We use $upSum[i]$ to denote the total weight of spanning rectangles that span slab $\gamma_i$ and currently intersect the sweeping line; $upSum[i]$ is initially set to 0 (Line 2). Also, we set $t_{slab}[i]$ to be the tuple representing the max-interval of $\gamma_i$ in the sweeping line. Since we sweep the line from bottom to top, we initially set $t_{slab}[i].y = -\infty$. In addition, the initial interval and sum of $t_{slab}[i]$ are set to be the x-range of $\gamma_i$ and 0, respectively (Line 3). When the sweeping line encounters the bottom of a spanning rectangle that spans $\gamma_i$, we add the weight of the rectangle to $upSum[i]$ (Lines 6 - 8); conversely, when the sweeping line encounters the top of the spanning rectangle, we subtract the weight of the rectangle (Lines 9 - 11). When the sweeping line encounters several tuples (from different slab-files) having the same y-coordinate (Line 12), we first update $t_{slab}[i]$'s accordingly (Lines 13 - 16), and then identify the tuples with the maximum *sum* among all the $t_{slab}[i]$'s (Line 17). Since there can be multiple tuples with the same maximum *sum* at an h-line, we call a function *GetMaxInterval* to generate a single tuple from those tuples (Line 18). Specifically, given a set of tuples with the same *sum* value, GetMaxInterval simply performs:

1. If the max-intervals of some of those tuples are consecutive, merge them into one tuple with an extended max-interval.

2. Return an arbitrary one of the remaining tuples after the above step.

Lastly, we insert the tuple generated from GetMaxInterval into the slab-file to be returned (Line 20). This process will continue until the sweeping line reaches the end of all the slab files and the set of spanning rectangles.

---

**Algorithm 1** MergeSweep

**Input:** $m$ slab-files $S_1, ..., S_m$ for $m$ slabs $\gamma_1, ..., \gamma_m$, a set of spanning rectangles $R'$

**Output:** a slab-file $S$ for slab $\gamma = \bigcup_{i=1}^{m} \gamma_i$. Initially $S \leftarrow \phi$

1: **for** $i = 0$ to $m$ **do**
2:    $upSum[i] \leftarrow 0$
3:    $t_{slab}[i] \leftarrow\ <-\infty,$ the range of x-coordinates of $\gamma_i, 0 >$
4: **end for**
5: **while** sweeping the horizontal line $\ell$ from bottom to top **do**
6:    **if** $\ell$ meets the bottom of $r_o \in R'$ **then**
7:      $upSum[j] \leftarrow upSum[j] + w(o), \forall j$ s.t. $r_o$ spans $\gamma_j$
8:    **end if**
9:    **if** $\ell$ meets the top of $r_o \in R'$ **then**
10:     $upSum[j] \leftarrow upSum[j] - w(o), \forall j$ s.t. $r_o$ spans $\gamma_j$
11:    **end if**
12:    **if** $\ell$ meets a set of tuples $T = \{t \mid t.y = \ell.y\}$ **then**
13:     **for all** $t \in T$ **do**
14:       $t_{slab}[i] \leftarrow t$, s.t. $t \in S_i$
15:       $t_{slab}[i].sum \leftarrow t.sum + upSum[i]$, s.t. $t \in S_i$
16:     **end for**
17:     $T' \leftarrow$ the set of tuples in $t_{slab}[1], ..., t_{slab}[m]$ with the largest *sum* values
18:     $t_{max} \leftarrow$ GetMaxInterval($T'$)
19:    **end if**
20:    $S \leftarrow S \cup \{t_{max}\}$
21: **end while**
22: **return** $S$

*Example 2.* Figure 7 shows how the MergeSweep algorithm works by using Example 1. For clarity, rectangles are removed, and the *sum* value of each max-interval is given above the segment representing the max-interval. Also, the value of *upSum* for each slab is given as a number enclosed in a bracket, e.g., $upSum[2] = 1$, between $y_2$ and $y_6$.

When the sweeping line $\ell$ is located at $y_0$, two max-intervals from $\gamma_3$ and $\gamma_4$ are merged into a larger max-interval. On the other hand, when $\ell$ is located at $y_1$, the max-interval from $\gamma_4$ is chosen, since its *sum* value 2 is the maximum among the 2 max-intervals at $y_1$. In addition, it is important to note that *sum* values of the max-intervals at $y_4$ and $y_5$ are increased by the value of $upSum[2] = 1$. Figure 7(b) shows the resulting max-intervals at the end of merging slab-files. We can find that the max-region of the entire data space is between max-intervals at $y_4$ and $y_5$, because the max-interval at $y_4$ has the highest *sum* value 3.

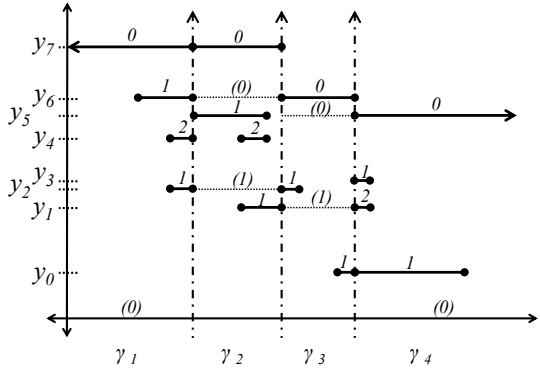

(a) Four slab-files before merge

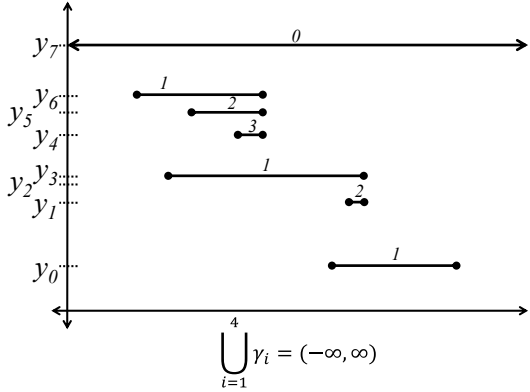

(b) A slab-file after merge

**Figure 7: An example to illustrate MergeSweep algorithm**

We can derive the following lemma:

LEMMA 3. *Let $K$ be the number of rectangles in slab $\gamma$ in a certain recursion. Given $m$ slab-files $S_1,...,S_m$ of slabs $\gamma_1,...,\gamma_m$, s.t., $\gamma = \cup_{i=1}^m \gamma_i$, and a set of spanning rectangles $R'$, MergeSweep algorithm returns the slab-file $S$ of $\gamma$ in $O(K/B)$ I/O's.*

PROOF. Since we set $m = \Theta(M/B)$, a block of memory can be allocated as the input buffer for each slab-file as well as the file containing spanning rectangles. Also, we use another block of memory for the output buffer. By doing this, we can read a tuple of slab-files or a spanning rectangle, or write a tuple to the merged slab-file in $O(1/B)$ I/O's amortized.

The number of I/O's performed by MergeSweep is proportional to the total number of tuples of all slab-files plus the number of spanning rectangles, i.e., $O((|R'|+\sum_{i=1}^m |S_i|)/B)$. Let $K_i$ be the number of rectangles in $\gamma_i$. Then $|S_i| = O(K_i)$ by Lemma 2. Also, $K_i = \Theta(K/m)$, since the $2K$ vertical edges of the $K$ rectangles are divided into $m$ slabs evenly. Therefore, $\sum_{i=1}^m |S_i| = O(K)$, which leads $O((|R'|+\sum_{i=1}^m |S_i|)/B) = O(K/B)$, since $|R'| \leq K$. □

### 5.2.4 Overall Algorithm

The overall recursive algorithm ExactMaxRS is presented in Algorithm 2. We can obtain the final slab-file with regard to a set $R$ of rectangles by calling ExactMaxRS($R$, $\gamma$, $m$), where the x-range of $\gamma$ is $(-\infty,\infty)$. Note that when the input set of rectangles can fit in the main memory, we invoke *PlaneSweep(R)* (Line 9), which is an in-memory algorithm that does not cause any I/O's.

---
**Algorithm 2** ExactMaxRS
---
**Input:** a set of rectangles $R$, a slab $\gamma$, the number of sub-slabs $m$
**Output:** a slab-file $S$ for $\gamma$

1: **if** $|R| > M$ **then**
2:     Partition $\gamma$ into $\gamma_1,...,\gamma_m$, which have roughly the same number of rectangles.
3:     Divide $R$ into $R_1,...,R_m$, $R'$, where $R_i$ is the set of non-spanning rectangles whose left (or right) vertical edges are in $\gamma_i$ and $R'$ is the set of spanning rectangles.
4:     **for** $i = 1$ to $m$ **do**
5:         $S_i \leftarrow$ ExactMaxRS($R_i$, $\gamma_i$, $m$)
6:     **end for**
7:     $S \leftarrow$ MergeSweep($S_1,...,S_m$, $R'$)
8: **else**
9:     $S \leftarrow$ PlaneSweep($R$)
10: **end if**
11: **return** $S$

---

From returned $S$, we can find the max-region by comparing *sum* values of tuples trivially. After finding the max-region, an optimal point for the MaxRS problem can be any point in the max-region, as mentioned in Section 5.1.

The correctness of Algorithm 2 is proved by the following lemma and theorem:

LEMMA 4. *Let $I^*$ be a max-interval at a h-line with regard to the entire space and $I_1^*,...,I_\mu^*$ be consecutive pieces of $I^*$ for a recursion, each of which belongs to slab $\gamma_i$, where $1 \leq i \leq \mu$. Then $I_i^*$ is also the max-interval at the h-line with regard to slab $\gamma_i$.*

PROOF. Let $sum(I)$ be the *sum* value of interval $I$. To prove the lemma by contradiction, suppose that there exists $I_i^*$ that is not a max-interval in $\gamma_i$. Thus, there exists $I'$ in $\gamma_i$ such that $sum(I') > sum(I_i^*)$ on the same h-line. For any upper level of recursion, if no rectangle spans $\gamma_i$, then $sum(I')$ and $sum(I_i^*)$ themselves are already the sum values with regard to the entire space. On the contrary, if there exist rectangles that span $\gamma_i$ at some upper level of recursion, then the sum values of $I'$ and $I_i^*$ with regard to the entire space will be $sum(I') + W_{span}$ and $sum(I_i^*) + W_{span}$, where $W_{span}$ is the total sum of the weights of all the rectangles spanning $\gamma_i$ in all the upper level of recursion. In both cases above, $sum(I') > sum(I_i^*)$ with regard to the entire space, which contradicts that $I^*$ is the max-interval with regard to the entire space. □

THEOREM 1. *The slab-file returned from the ExactMaxRS algorithm is correct with regard to a given dataset.*

PROOF. Let $\rho^*$ be the max-region with regard to a given dataset, and similarly $I^*$ be the best max-interval that is in fact the bottom edge of $\rho^*$. Then we want to prove that the algorithm eventually returns a slab-file which contains $I^*$.

Also, by Lemma 4, we can claim that for any level of recursion, a component interval $I_i^*$ of $I^*$ will also be the max-interval for its h-line within slab $\gamma_i$. By Algorithm 1, for each h-line, the best one among the max-intervals at each h-line is selected (perhaps also extended). Therefore, eventually $I^*$ will be selected as a max-interval with regard to the entire space. □

Moreover, we can prove the I/O-efficiency of the Exact-MaxRS algorithm as in the following theorem:

THEOREM 2. *The ExactMaxRS algorithm solves the MaxRS problem in $O((N/B) \log_{M/B} (N/B))$ I/O's, which is optimal in the EM model among all comparison-based algorithms.*

PROOF. The dataset needs to be sorted by x-coordinates before it is fed into Algorithm 2. The sorting can be done in $O((N/B) \log_{M/B} (N/B))$ I/O's using the textbook-algorithm external sort.

Given a dataset with cardinality $N$ sorted by x-coordinates, the decomposition of the dataset along the x-dimension can be performed in linear time, i.e., $O(N/B)$. Also, by Lemma 3, the total I/O cost of the merging process at each recursion level is also $O(N/B)$, since there can be at most $2N$ rectangles in the input of any recursion. By the proof of Lemma 1, there are $O(\log_{M/B} (N/B))$ levels of recursion. Hence, the total I/O cost is $O((N/B) \log_{M/B} (N/B))$.

The optimality of this I/O complexity follows directly from the results of [2] and [11]. □

## 6. APPROXIMATION ALGORITHM FOR MAXIMIZING CIRCULAR RANGE SUM

In this section, we propose an approximation algorithm, namely *ApproxMaxCRS*, for solving the MaxCRS problem (Definition 2). Our algorithm finds an (1/4)-approximate solution in $O((N/B) \log_{M/B} (N/B))$ I/O's. We achieve the purpose by a novel reduction that converts the MaxCRS problem to the MaxRS problem.

### 6.1 ApproxMaxCRS

Recall (from Definition 2) that the goal of the MaxCRS problem is to find a circle with a designated diameter that maximizes the total weight of the points covered. Denote by $d$ the diameter. Following the idea explained in Section 4, first we transform the MaxCRS problem into the following problem: Let $C$ be a set of circles each of which is centered at a distinct object $o \in O$, has a diameter as specified in the MaxCRS problem, and carries a weight $w(o)$. We want to find a location $p$ in the data space to maximize the total weight of the circles in $C$ covering $p$. Figure 8(a) shows an instance of the transformed MaxCRS problem, where there are four circles in $C$, each of which is centered at an object $o \in O$ in the original MaxCRS problem. An optimal answer can be any point in the gray area.

We will use the ExactMaxRS algorithm developed in the previous section as a tool to compute a good approximate answer for the MaxCRS problem. For this purpose, we convert each circle of $C$ to its Minimum Bounding Rectangle (MBR). Obviously, the MBR is a $d \times d$ square. Let $R$ be the set of resulting MBRs. Now, apply ExactMaxRS on $R$, which outputs the max-region with regard to $R$. Understandably, the max-region (black area in Figure 8(b)) returned from the ExactMaxRS algorithm may contain locations that are suboptimal for the original MaxCRS problem (in Figure 8(b), only points in the gray area are optimal). Moreover, in the worst case, the max-region may not even intersect with any circle at all as shown in Figure 8(c).

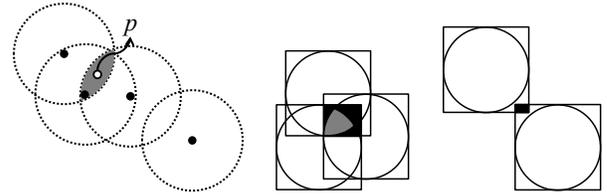

(a) The transformed MaxCRS problem (b) MBRs of circles (c) Worst case

**Figure 8: Converting MaxCRS to MaxRS**

Therefore, in order to guarantee the approximation bound, it is insufficient to just return a point in the max region. Instead, our ApproxMaxCRS algorithm returns the best point among the center of the max-region and four *shifted point*s. The algorithm is presented in Algorithm 3.

---
**Algorithm 3** ApproxMaxCRS
---
**Input:** a set of circles $C$, a slab $\gamma$ whose range of the x-coordinate is $(-\infty, \infty)$, the number of slabs $m$
**Output:** a point $\hat{p}$

1: Construct a set $R$ of MBRs from $C$
2: $\rho \leftarrow$ ExactMaxRS$(R, \gamma, m)$
3: $p_0 \leftarrow$ the center point of $\rho$
4: **for** $i = 1$ to $4$ **do**
5:     $p_i \leftarrow$ GetShiftedPoint$(p_0, i)$
6: **end for**
7: $\hat{p} \leftarrow$ the point $p$ among $p_0, ..., p_4$ that maximizes the total weight of the circles covering $p$
8: **return** $\hat{p}$
---

After obtaining the center point $p_0$ of the max-region $\rho$ returned from ExactMaxRS function (Lines 2 - 3), we find four shifted points $p_i$, where $1 \leq i \leq 4$, from $p_0$ as shown in Figure 9 (Lines 4 - 6). We use $\sigma$ to denote the *shifting distance* which determines how far a shifted point should be away from the center point. To guarantee the approximation bound as proved in Section 6.2, $\sigma$ can be set to any value

1095

| Symbol | Description |
| --- | --- |
| $d$ | the diameter of circles (a given parameter of the MaxCRS problem) |
| $p_0$ | the centroid of the max-region returned by ExactMaxRS |
| $p_i$ ($i \in [1, 4]$) | a shifted point described in Algorithm 3 |
| $c_i$ ($i \in [0, 4]$) | the circle with diameter $d$ centering at point $p_i$ |
| $r_0$ | the MBR of $c_0$ |
| $O(s)$ | the set of objects covered by $s$, where $s$ is a circle or an MBR |
| $W(s)$ | the total weight of the objects in $O(s)$ |

**Table 1: List of notations**

such that $(\sqrt{2}-1)\frac{d}{2} < \sigma < \frac{d}{2}$. Finally, we return the best point $\hat{p}$ among $p_0, ..., p_4$ (Lines 7 - 8).

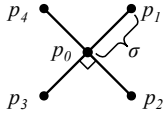

**Figure 9: The illustration of shifting points**

Note that Algorithm 3 does not change the I/O complexity of the ExactMaxRS algorithm, since only linear I/O's are required in the entire process other than running the ExactMaxRS algorithm. Note that Line 7 of Algorithm 3 requires only a single scan of $C$.

## 6.2 Approximation Bound

Now, we prove that the ApproxMaxCRS algorithm returns a (1/4)-approximate answer to the optimal solution, and also prove that this approximation ratio is tight with regard to this algorithm. To prove the approximation bound, we use the fact that a point $p$ covered by the set of circles (or MBRs) in the transformed MaxCRS problem is truly the point such that the circle (or MBR) centered at $p$ covers the corresponding set of objects in the original MaxCRS problem. The main notations used in this section are summarized in Table 1.

LEMMA 5. *For each $i \in [0, 4]$, let $c_i$ be the circle centered at point $p_i$, $r_0$ be the MBR of $c_0$, and $O(s)$ be the set of objects covered by $s$, where $s$ is a circle or an MBR. Then $O(r_0) \subseteq O(c_1) \cup O(c_2) \cup O(c_3) \cup O(c_4)$.*

PROOF. As shown in Figure 10, all the objects covered by $r_0$ are also covered by $c_1$, $c_2$, $c_3$, or $c_4$, since $(\sqrt{2}-1)\frac{d}{2} < \sigma < \frac{d}{2}$. □

Let $W(s)$ be the total weight of the objects covered by $s$, where $s$ is a circle or an MBR. Then, we have:

LEMMA 6. $W(r_0) \leq 4 \max_{0 \leq i \leq 4} W(c_i)$.

PROOF.

$$\begin{aligned} W(r_0) &\leq \sum_{1 \leq i \leq 4} W(c_i) \quad \text{(by Lemma 5)} \\ &\leq 4 \max_{0 \leq i \leq 4} W(c_i) \end{aligned}$$

□

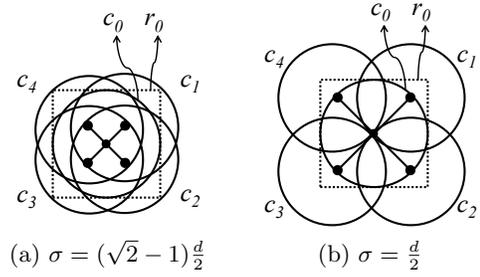

**Figure 10: Lemma 5**

THEOREM 3. *The ApproxMaxCRS algorithm returns a (1/4)-approximate answer to the MaxCRS problem.*

PROOF. Recall that $\hat{p}$ is the point returned from Algorithm 3 as the approximate answer to the MaxCRS problem. Let point $p^*$ be an optimal answer for the MaxCRS problem. Denote by $\hat{r}$ and $r^*$ the MBRs centered at point $\hat{p}$ and $p^*$, respectively. Likewise, denote by $\hat{c}$ and $c^*$ be the circles centered at point $\hat{p}$ and $p^*$, respectively. The goal is to prove $W(c^*) \leq 4W(\hat{c})$.

We achieve this purpose with the following derivation:

$$W(c^*) \leq W(r^*) \leq W(r_0) \leq 4 \max_{0 \leq i \leq 4} W(c_i) = 4W(\hat{c})$$

The first inequality is because $r^*$ is the MBR of $c^*$. The second inequality is because $p_0$ is the optimal solution for the MaxRS problem on $R$. The last equality is because ApproxMaxCRS returns the best point among $p_0, ..., p_4$. □

THEOREM 4. *The 1/4 approximation ratio is tight for the ApproxMaxCRS algorithm.*

PROOF. We prove this by giving a worst case example. Consider an instance of the transformed MaxCRS problem in Figure 11 where each circle has weight 1. In this case, we may end up finding a max-region centered at $p_0$ using the ExactMaxRS algorithm (notice that both $p_0$ and $p^*$ are covered by 4 MBRs). In this case, we will choose one of $p_1, ..., p_4$ as an approximate solution. Since each of $p_1, ..., p_4$ is covered by only 1 circle, our answer is (1/4)-approximate, because the optimal answer $p^*$ is covered by 4 circles. □

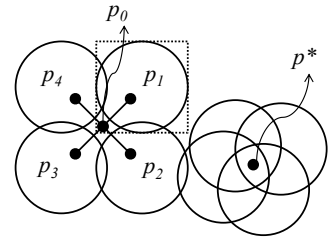

**Figure 11: Theorem 4**

## 7. EMPIRICAL STUDY

In this section, we evaluate the performance of our algorithms with extensive experiments.

| Dataset | Cardinality |
|---------|-------------|
| UX | 19,499 |
| NE | 123,593 |

Table 2: The cardinalities of real datasets

| Parameter | Default value |
|-----------|---------------|
| Cardinality ($|O|$) | 250,000 |
| Block size | 4KB |
| Buffer size | 256KB (real dataset), 1024KB (synthetic dataset) |
| Space size | $1M \times 1M$ |
| Rectangle size ($d_1 \times d_2$) | $1K \times 1K$ |
| Circle diameter ($d$) | $1K$ |

Table 3: The default values of parameters

## 7.1 Environment Setting

We use both real and synthetic datasets in the experiments. We first generate synthetic datasets under uniform distribution and Gaussian distribution. We set the cardinalities of dataset (i.e., $|O|$) to be from 100,000 to 500,000 (default 250,000). The range of each coordinate is set to be $[0, 4|O|]$ (default $[0, 1000000]$).

We also use two real datasets, *North East (NE) dataset* and *United States of America and Mexico (UX) dataset*, downloaded from the R-tree Portal [1]. The cardinalities of datasets are presented in Table 2. For both datasets, we normalize the range of coordinates to $[0, 1000000]$.

Since no method is directly applicable to the MaxRS problem in spatial databases, we should externalize the in-memory algorithm [11, 14] for max-rectangle enclosing problem to be compared with our ExactMaxRS algorithm. In fact, the externalization of this in-memory algorithm is already proposed by Du et al. [9], which is originally invented for processing their optimal-location queries. They present two algorithms based on plane-sweep, called *Naive Plane Sweep* and *aSB-Tree*, which are also applicable to the MaxRS problem, even though their main algorithm based on a preprocessed structure, called the *Vol-Tree*, cannot be used in the MaxRS problem.

As a performance metric, we use the number of I/O's, precisely the number of transferred blocks during the entire process. We do not consider CPU time, since it is dominated by I/O cost.

We fix the block size to 4KB, and set the buffer size to 256KB for real datasets and 1024KB for synthetic datasets by default. This is because the cardinalities of the real datasets are relatively small (recall that we consider a massive dataset which cannot be fully loaded into the main memory). Also, for the MaxRS problem, we set the rectangle size to $1000 \times 1000$ by default. Similarly, for the MaxCRS problem, we set the circle diameter to 1000 by default. All the default values of parameters are presented in Table 3.

We implement all the algorithms in Java, and conduct all the experiments on a PC equipped with Intel Core i7 CPU 3.4GHz and 16GB memory.

## 7.2 Experimental Results

In this section, we present our experimental results. First, we examine the performance of alternative algorithms in terms of I/O cost by varying the parameters. Note that the I/O cost is in log scale in all the relevant graphs. We finally show the quality of approximation of our ApproxMaxCRS algorithm in Section 7.2.5.

### 7.2.1 Effect of the Dataset Cardinalities

Figure 12 shows the experimental results for varying the total number of objects in the dataset. Both of the results of Gaussian distribution and uniform distribution shows our ExactMaxRS is much more efficient than the algorithms based on plane-sweep. Especially, even if the dataset gets larger, the ExactMaxRS algorithm achieves performance similar to that on the smaller dataset, which effectively shows that our algorithm is scalable to datasets of a massive size.

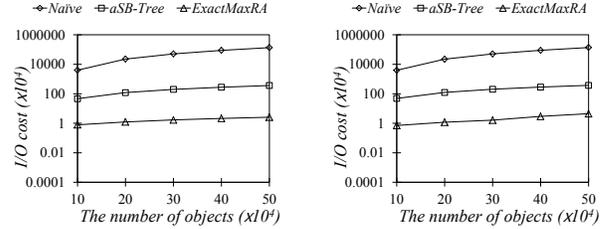

(a) Gaussian distribution  (b) Uniform distribution

Figure 12: Effect of the dataset cardinalities

### 7.2.2 Effect of the Buffer Size

Figure 13 shows the experimental results for varying the buffer size. Even though all the algorithms exhibit better performance as the buffer size increases, the ExactMaxRS algorithm is more sensitive to the size of buffer than the others. This is because our algorithm uses the buffer more effectively. As proved in Theorem 2, the I/O complexity of ExactMaxRS is $O((N/B) \log_{M/B}(N/B))$, which means the larger $M$, the smaller the factor $\log_{M/B}(N/B)$. Nevertheless, once the buffer size is larger than a certain size, the ExactMaxRS algorithm also shows behavior similar to the others, since the entire I/O cost will be dominated by $O(N/B)$, i.e., linear pcost.

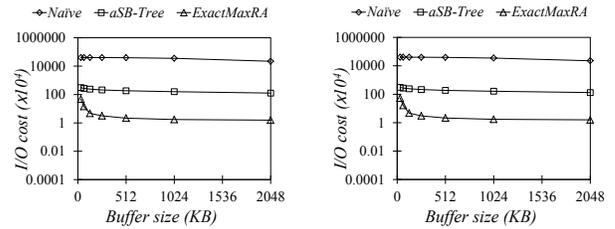

(a) Gaussian distribution  (b) Uniform distribution

Figure 13: Effect of the buffer size

### 7.2.3 Effect of the Range Size

Figure 14 shows the experimental results for varying the range parameters. Without loss of generality, we use the same value for each dimension, i.e., each rectangle is a square. It is observed that the ExactMaxRS algorithm is less influenced by the size of range than the other algorithms. This is because as the size of range increases, the probability that rectangles overlap also increases in the algorithms based on plane-sweep, which means that the number of interval insertions will also increase. Meanwhile, the ExactMaxRS algorithm is not much affected by the overlapping probability.

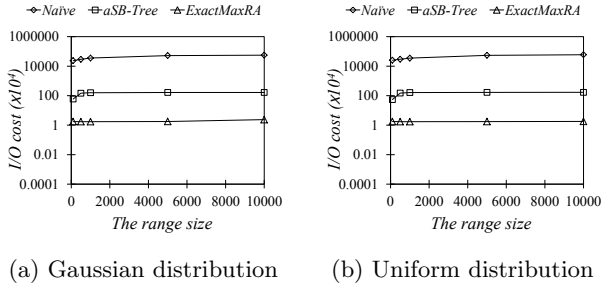

(a) Gaussian distribution     (b) Uniform distribution

**Figure 14: Effect of the range size**

### 7.2.4 Results of Real Datasets

We conduct the same kind of experiments on real datasets except varying cardinalities. As shown in Table 2, dataset UX is not only much smaller, but also sparser than NE, since the domains of the data space are the same, i.e., $1M \times 1M$. In fact, we can regard UX as a macro view of NE.

Overall trends of the graphs are similar to the results in synthetic datasets, as shown in Figures 15 and 16. Note that in Figure 15(a), when the buffer size gets larger than 512KB, the naive plane sweep algorithm shows the best performance. This is because UX is small enough to be loaded into a buffer of size 512KB, which causes only one linear scan. However, we can see that the aSB-Tree cannot be loaded into a buffer of the same size, since the aSB-Tree requires more space due to the other information in the tree structure such as pointers of child nodes.

In this paper, since we focus on massive datasets that should be stored in external memory, we can claim that our ExactMaxRS algorithm is much more efficient than the others for large datasets such as NE.

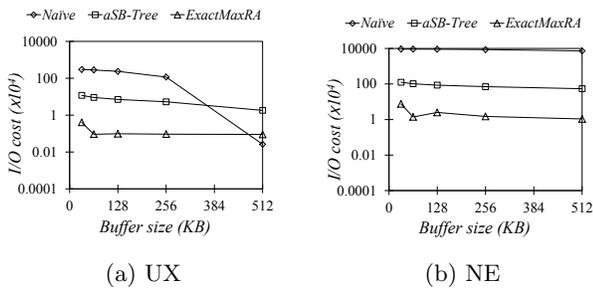

(a) UX     (b) NE

**Figure 15: Effect of the buffer size on real datasets**

### 7.2.5 The Quality of Approximation

Finally, we evaluate the quality of approximation obtained from the ApproxMaxCRS algorithm in Figure 17. Since the quality can be different when the diameter $d$ changes, we examine the quality by varying $d$ on both synthetic and real datasets. Optimal answers are obtained by implementing a theoretical algorithm [8] that has time complexity $O(n^2 \log n)$ (and therefore, is not practical). We observe that when the diameter gets larger, the quality of approximation becomes higher and more stable, since more objects are included in the given range. Even though theoretically our ApproxMaxCRS algorithm guarantees the (1/4)-approximation bound, the average approximation ratio is much larger than 1/4 in practice, which is close to 0.9.

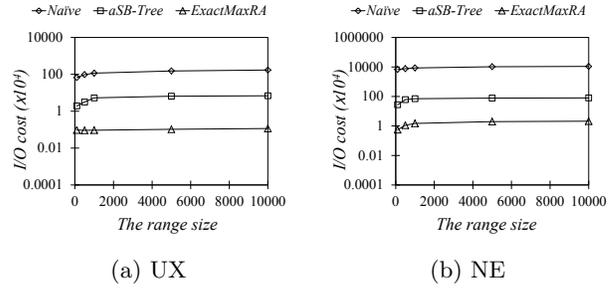

(a) UX     (b) NE

**Figure 16: Effect of the range size on real datasets**

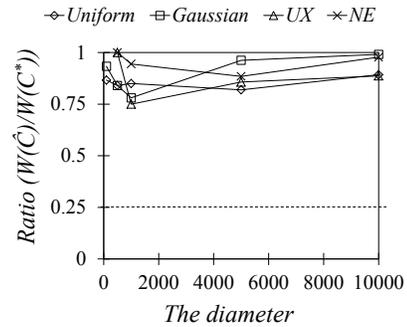

**Figure 17: Approximation quality**

## 8. CONCLUSIONS AND FUTURE WORK

In this paper, we solve the MaxRS problem in spatial databases. This problem is useful in many scenarios such as finding the most profitable service place and finding the most serviceable place, where a certain size of range should be associated with the place. For the MaxRS problem, we propose the first external-memory algorithm, ExactMaxRS, with a proof that the ExactMaxRS algorithm correctly solves the MaxRS problem in optimal I/O's. Furthermore, we propose an approximation algorithm, ApproxMaxCRS, for the MaxCRS problem that is a circle version of the MaxRS problem. We also prove that the ApproxMaxCRS algorithm gives a (1/4)-approximate solution to the exact solution for the MaxCRS problem. Through extensive experiments on both synthetic and real datasets, we demonstrate that the proposed algorithms are also efficient in practice.

Now we are considering several directions for our future works. First, it will be naturally feasible to extend our algorithm to deal with Max$k$RS problem or *MinRS* problem. Second, focusing on the MaxCRS problem, we are planning to improve the algorithm to give a tighter bound. Finally, although our ExactMaxRS algorithm is proved optimal in terms of I/O cost, so far we do not use any preprocessed structure. Therefore, our next direction can be to reduce the searching cost by using a newly invented index.

## 9. ACKNOWLEDGMENTS

This work was supported in part by WCU (World Class University) program under the National Research Foundation of Korea funded by the Ministry of Education, Science and Technology of Korea (No. R31-30007), and in part by the National Research Foundation of Korea grant funded by the Korean government (MEST) (No. 2012-0000182).

Yufei Tao was supported in part by project GRF 4165/11 from HKRGC.